\documentclass[final,5p,times,twocolumn]{elsarticle}

\usepackage{graphicx}
\usepackage{amssymb}
\usepackage{amsmath}
\usepackage{comment}
\usepackage[colorlinks=true,linkcolor=NavyBlue,citecolor=NavyBlue,urlcolor=NavyBlue]{hyperref}

\biboptions{sort&compress} 

\makeatletter
\def\@author#1{\g@addto@macro\elsauthors{\normalsize%
    \def\baselinestretch{1}%
    \upshape\authorsep#1\unskip\textsuperscript{%
      \ifx\@fnmark\@empty\else\unskip\sep\@fnmark\let\sep=,\fi
      \ifx\@corref\@empty\else\unskip\sep\@corref\let\sep=,\fi
      }%
    \def\authorsep{\unskip,\space}%
    \global\let\@fnmark\@empty
    \global\let\@corref\@empty  
    \global\let\sep\@empty}%
    \@eadauthor={#1}
}
\makeatother

\makeatletter
\def\ps@pprintTitle{%
 \let\@oddhead\@empty 
 \let\@evenhead\@empty
 \def\@oddfoot{}%
 \let\@evenfoot\@oddfoot}
 \makeatother

\journal{Physical Review E}
\begin{document}
\begin{frontmatter}

\title{Currency target zone modeling: An interplay between physics and economics}

\author[add1]{Sandro Claudio Lera}
\ead{slera@ethz.ch}

\author[add2,add3]{Didier Sornette}
\ead{dsornette@ethz.ch}

\address[add1]{ETH Zurich, Singapore-ETH Centre, 1 CREATE Way, \#06-01 CREATE Tower, 138602 Singapore}
\address[add2]{ETH Zurich, Department of Management, Technology, and Economics, Scheuchzerstrasse 7, 8092 Zurich, Switzerland}
\address[add3]{Swiss Finance Institute, c/o University of Geneva, Geneva, Switzerland}

\begin{abstract}

We study the performance of the euro/Swiss franc exchange rate
in the extraordinary period from September 6, 2011 and January 15, 2015 when the 
Swiss National Bank enforced a minimum exchange 
rate of 1.20 Swiss francs per euro. Within the general framework 
built on geometric Brownian motions (GBM) and based on the analogy between
Brownian motion in finance and physics, the first-order
effect of such a steric constraint would enter a priori in the form of a repulsive entropic force
associated with the paths crossing the barrier that are forbidden. Non-parametric empirical
estimates of drift and volatility show that the predicted first-order analogy 
between economics and physics are incorrect. 
The clue is to realise that the random walk nature of financial prices results from
the continuous anticipations of traders about future opportunities, whose aggregate
actions translate into an approximate efficient market with almost no arbitrage opportunities.
With the Swiss National Bank stated commitment to enforce the barrier,
traders's anticipation of this action leads to a vanishing drift together with a volatility of the exchange rate
that depends on the distance to the barrier. This effect 
is described by Krugman's model 
[P.R. Krugman. \textit{Target zones and exchange rate dynamics}. The Quarterly Journal of Economics, 106(3):669-682, 1991].
We give direct quantitative empirical evidence that Krugman's theoretical model provides
an accurate description of the euro/Swiss franc target zone. Motivated by the insights from the economical model, 
we revise the initial economics-physics analogy and show that, within the context of hindered diffusion, 
the two systems can be described with the same mathematics after all. 
Using a recently proposed extended analogy in terms of a colloidal Brownian particle embedded in a fluid of molecules
associated with the underlying order book, we derive that, close to the
restricting boundary, the dynamics of both systems is described by a stochastic
differential equation with a very small constant drift and a linear diffusion coefficient. 
As a side result, we present a simplified derivation of the linear hydrodynamic diffusion coefficient of 
a Brownian particle close to a wall. 
\end{abstract}

\begin{keyword}
Exchange rate dynamics \sep target zone \sep order book fluid \sep econophysics
\PACS 89.65.Gh, 05.40.Jc,89.75.-k
\end{keyword}
\end{frontmatter}

\section{Introduction}
\label{sec:Introduction}

Perhaps not apparent at first glance, physics and economics have been life-long companions during their mutual 
development of concepts and methods emerging in both fields. 
There has been much mutual enrichment and cross-fertilization. Since the beginning of the formulation of the scientific 
approach in the physical and natural sciences, economists have taken inspiration from physics, in particular in its success in describing 
natural regularities and processes \cite{Sornette2014}. Already in 1776, Adam Smith formulated his ``Inquiry into the Nature
and Causes of the Wealth of Nations" inspired by the ``Philosophiae Naturals Principia Mathematica" (1687) of 
Sir Isaac Newton, which specifically stresses the notion of causative forces. 
Reciprocally, physics has been inspired several times by observations in economics. 
A prominent example of this kind is the theory of Brownian motion and random walks. 
In order to model the apparent random walk motion of bonds and stock options in the Paris stock market, mathematician Louis Bachelier \cite{Bachelier1900} 
developed in his thesis the mathematical theory of diffusion. 
He solved the parabolic diffusion equation five years before Albert Einstein \cite{Einstein1905} 
established the theory of Brownian motion based on the same diffusion equation, also underpinning the theory of random walks. 
These two works have ushered research on mathematical descriptions of fluctuation phenomena in statistical physics, 
of quantum fluctuation processes in elementary particles-fields physics, on the one hand, and of financial prices on the other hand, 
both anchored in the random walk model and Wiener process. \\ 
Here, we will extend this analogy even further and describe a restricted Brownian motion 
both from a physical and an economical perspective, paying close attention to the interplay between the two fields. 
The purpose of this paper is thereby twofold. On the one hand, we provide novel evidence to support
the famous, yet often times refuted Krugman target zone model. Since this aspect is, however, 
primarily of interest for the economics community,  a more detailed report of these results
will be published elsewhere \cite{Lera2015a}.
More importantly, here, we use the example of a restricted Brownian motion
to point out several new observations made about the interplay between natural and social 
sciences, which are relevant in particular for the physicist aspiring to do interdisciplinary research 
in economics and finance.

\section{Target zone arrangements}
\label{sec:tz_arrangements}

\begin{figure*}
	\centering
	\includegraphics[width = \textwidth]{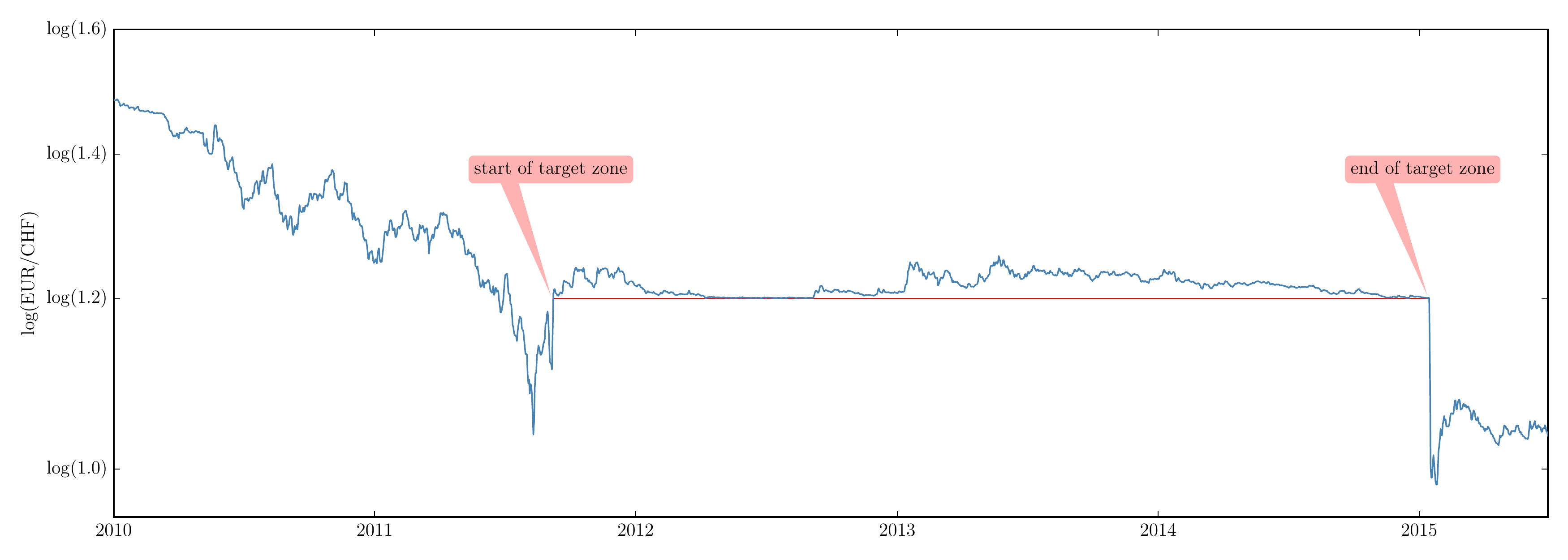}
	\caption{We show the Euro/Swiss franc (EUR/CHF) exchange rate between January 1, 2010 and  June 30, 2015. On September 6, 2011, 
	the Swiss National Bank (SNB) officially announced its decision to enforce a minimum of $1.20$ Swiss francs per euro by buying 
	euros and selling Swiss francs in unlimited amounts if necessary.}
	\label{fig:EURCHF}
\end{figure*}

On September 6, 2011, the Swiss National Bank (SNB) announced that it would
enforce a minimum exchange rate of $1.20$ Swiss francs 
(CHF) per euro (EUR), in response to the European debt crisis and a continuously weakening euro. 
With a level around 1.6 CHF at the introduction of the euro in 1999 and a peak above 1.67 CHF on October 2007, 
the EUR/CHF has been floating freely until it dived to the record low of CHF 1.0070 per euro on August 9, 2011. 
The Swiss National Bank intervened massively leading to a fast rebound of the euro. On September 6, the SNB 
announced officially that it would defend the minimum exchange rate of CHF 1.20 by all means 
(buying euros and selling Swiss francs in unlimited amounts as deemed necessary). The SNB held this policy until 
January 15, 2015, leading to an exchange rate
levelling off between 1.20 and 1.24 CHF per euro, exhibiting a dynamics that is spectacularly different from what is 
observed for a freely floating currency pair, as can be seen in figure \ref{fig:EURCHF}. 

Such a restriction of an exchange rate is known in the finance literature as a target zone arrangement.
The central bank
announces that the exchange rate between the domestic and a foreign currency as determined by the free
market will not be let to exceed some pre-defined lower boundary (an upper boundary is analogous, two simultaneous
boundaries lead to slight modifications that are not further relevant for this discussion). 
As long as the exchange rate is well above the target zone boundary, the exchange rate is
allowed to fluctuate freely, controlled only by supply and demand on the foreign-exchange market (FX).
Once the exchange rate approaches the defended limit from above, the central bank intervenes by selling its domestic currency
to buy foreign exchange, thus increasing supply of the domestic and demand of the foreign currency. This pushes
the exchange rate back above its lower limit.  

The most general ansatz to model the dynamics of a financial time series is represented by
\begin{equation}
\frac{\mathrm{d}s}{\mathrm{d}t} = f(s,t) + g(s,t) \cdot \eta(t)
\label{eq:basic_SDE}
\end{equation}
where $s$ is the logarithm of the exchange rate (EUR/CHF exchange rate in this article),
with $\eta$ a Gaussian white noise and $f,g$ are two functions 
representing respectively the drift or expected return and the volatility (standard deviation). This 
ansatz is general since higher order derivatives would imply temporal correlations, thus violating
the no-arbitrage condition at the heart of the efficient market hypothesis (EMH)  \cite{Fama1970,Fama1991,Fama2014}. 
(Interestingly, it has been pointed out by several physicists that exactly the use of second order differential
equations can have its merits even in finance, see \cite{Bouchaud1998,Farmer2002,Ide2002}.) 

Dealing with exchange rates requires a priori to pay attention to the so-called triangular arbitrage \cite{Drozdz2010},
as occurs when an additional currency, say US dollar (USD) is included. We then have to consider
the relations between the triplet of exchange rates EUR/CHF, CHF/USD and USD/EUR. 
The triangular (no-) arbitrage condition is that the number of CHF for 1 EUR should be 
equal to the product of the number of CHF per 1 USD  times the number of USD per 1 EUR.
And a similar condition extends to cycles involving more currencies of arbitrary lengths.
In practice, deviation from triangular (no-)arbitrage conditions may occur only 
at small time scales and disappears very fast as a result of the action of traders
taking advantage of inconsistent cross-rates. Since our work is focused on the EUR/CHF exchange rate
close to a boundary imposed only on it and other currencies are not directly 
targeted by the action of the central bank, 
we neglect such influence. An a priori justification is that the statistical properties
found for exchange rate price series are in general similar if not undistinguishable from those of equities.
Our results support this simplifying assumption.

We follow the usual convention that the exchange rate denotes the amount of domestic currency that is needed to buy one unit of 
foreign currency. 
Pure Brownian motion is recovered for $f = 0$ and $g$ constant, whereas
pure geometric brownian motion (GBM) \cite{Bachelier1900,Merton1971} 
denotes the special case of  $f$ and $g$ being
constant, which embodies EMH  that financial markets incorporate information 
so effectively that the resulting price trajectory is akin to a random walk 
with no possible arbitrage. In order to respect causality for the correct calculation of investments performance, 
this stochastic equation is understood in the It\^o-sense. In the remainder of this paper, we will
investigate what is the predicted shape of $f$ and $g$ from different points of view. 

\section{Target zone modeling I: A first physicist's approach}
\label{sec:tzm_physical}

Starting from the structure \eqref{eq:basic_SDE}, 
we investigate the nature of the minimal ingredients needed to capture the abnormal dynamics 
observed in figure \ref{fig:EURCHF}. 
The aberrant trajectory of the EUR/CHF log-exchange rate $s(t)$ is clearly embodied by the visible existence
of the barrier at $s = \underline{s} \equiv \log(1.20)$ and the tendency for $s(t)$ to remain very close to it between September 2011 and January 2015. 
The simplest direct application of the GBM model to this situation is to assume that $s(t)$ continues to follow
a simple random walk but now constrained to remain above the impenetrable cap at  $\underline{s}$. 
Since such a situation is not intrinsic to finance, but could just as well correspond to a (one-dimensional) physical Brownian particle 
that is restricted by a wall at $\underline{s}$, we can use this analogy to employ well known mathematical tools from physics.
Putting a wall constraint on a random walk is known to 
induce an effective entropic force acting on the particle, resulting from the reduction of path configurations
by reflecting all random walks that would cross the wall \cite{Chandrasekhar1943,Redner2001}. In $1+1$
dimensions (one spatial dimension and one temporal dimension),
the corresponding entropic repulsive force can be shown to derive from an effective 
long-range entropic potential $V_\text{ENT} = C/(s-\underline{s})$, where $C>0$ is a constant
\cite{Fisher1984,Sornette1994}. Intuitively, this self-similar long-range potential is associated with the 
relationship between the average distance to the wall and the
long wavelengths of the random walks that are suppressed by the rigid impenetrable barrier. 

\begin{figure}
	\centering
	\includegraphics[width = 0.4\textwidth]{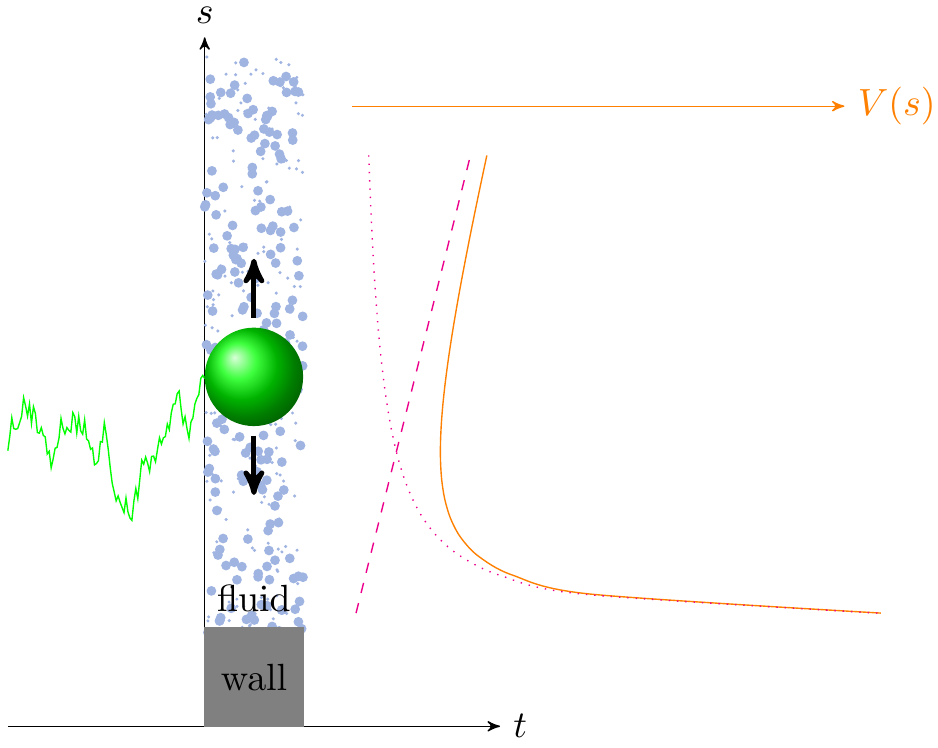}
	\caption{Random trajectory (the fluctuating, continuous line on the left) 
	of a one-dimensional Brownian particle moving in a potential $V(s)$ (continuous line on the right). 
	This potential is the sum of an attractive potential (dashed line) and a 
	repulsive potential (dotted line). From the most simplified physical perspective, 
	one would expect the EUR/CHF exchange rate
	between September 2011 and January 2015 to be controlled by such force potentials.}
	\label{fig:particle_in_potential}
\end{figure}

It is easy to see that the two ingredients  ``Brownian motion" and ``restricting wall" cannot account for 
the sustained proximity of the particle (exchange rate) to the wall (figure \ref{fig:EURCHF}). 
We need to add at least one more 
ingredient to account for this fact, which, in the economic picture, comes from the  strong economic 
``pressure" on the euro resulting from the European crisis, which led to the introduction of the $1.20$-cap in the first place. 
The simplest assumption is to assume a constant physical pressure that pushes the particle
towards the wall, corresponding to the linear potential $V_\text{ECO} = F \cdot (s-\underline{s})$ with a constant $F>0$.
Together, this yields the following total potential 
\begin{equation}
V \equiv  V_\text{ENT} + V_\text{ECO}  = \frac{C}{s-\underline{s}} +  F \cdot (s-\underline{s}),
\label{eq:potential} 
\end{equation}
depicted in figure \ref{fig:particle_in_potential}. 
The equilibrium position at which expression \eqref{eq:potential} finds its minimum
is $s_\text{eq} =\underline{s} + \sqrt{C/F}$: unsurprisingly, the stronger the pressure $F$ on the euro, the closer is the equilibrium
exchange rate to the barrier. Expanding \eqref{eq:potential}
around $s_\text{eq}$, using $f \equiv -\mathrm{d}V/\mathrm{d}s$ and inserting this into \eqref{eq:basic_SDE} gives to leading orders
\begin{equation}
\frac{\mathrm{d} s}{\mathrm{d}t} = 3 \frac{F^2}{C} \left(s - s_\text{eq} \right)^2 - 2 \sqrt{ \frac{F^3}{C}} \left(s - s_\text{eq} \right) + g \cdot \eta(t). 
\label{eq:physical_SDE}
\end{equation}
With equation \eqref{eq:physical_SDE},  we have derived a model aimed at capturing the constrained EUR/CHF dynamics
using only a minimal number of ingredients. Theoretically, 
one predicts from \eqref{eq:physical_SDE} a volatility scaling as $(s_\text{eq} -1.20)^{3/2}$ and a skewness scaling 
as $(s_\text{eq} -1.20)^{2}$, 
as can be derived without solving \eqref{eq:physical_SDE} using a path integral formalism and an
expansion in terms of Feynman diagrams \cite{Chow2010} (see \cite{Lera2015} for the detailed calculations).
One way to test the naive hypothesis \eqref{eq:physical_SDE} would be by calculating
the empirical moments from the data and comparing to the theoretical results. 
Instead, we choose a more direct test
and determine $f$ and $g$ empirically from the data.

\section{Empirical estimation of drift and volatility}
\label{sec:empirical_estimation}

\begin{figure}[!htb]
	\centering
	\includegraphics[width = 0.5\textwidth]{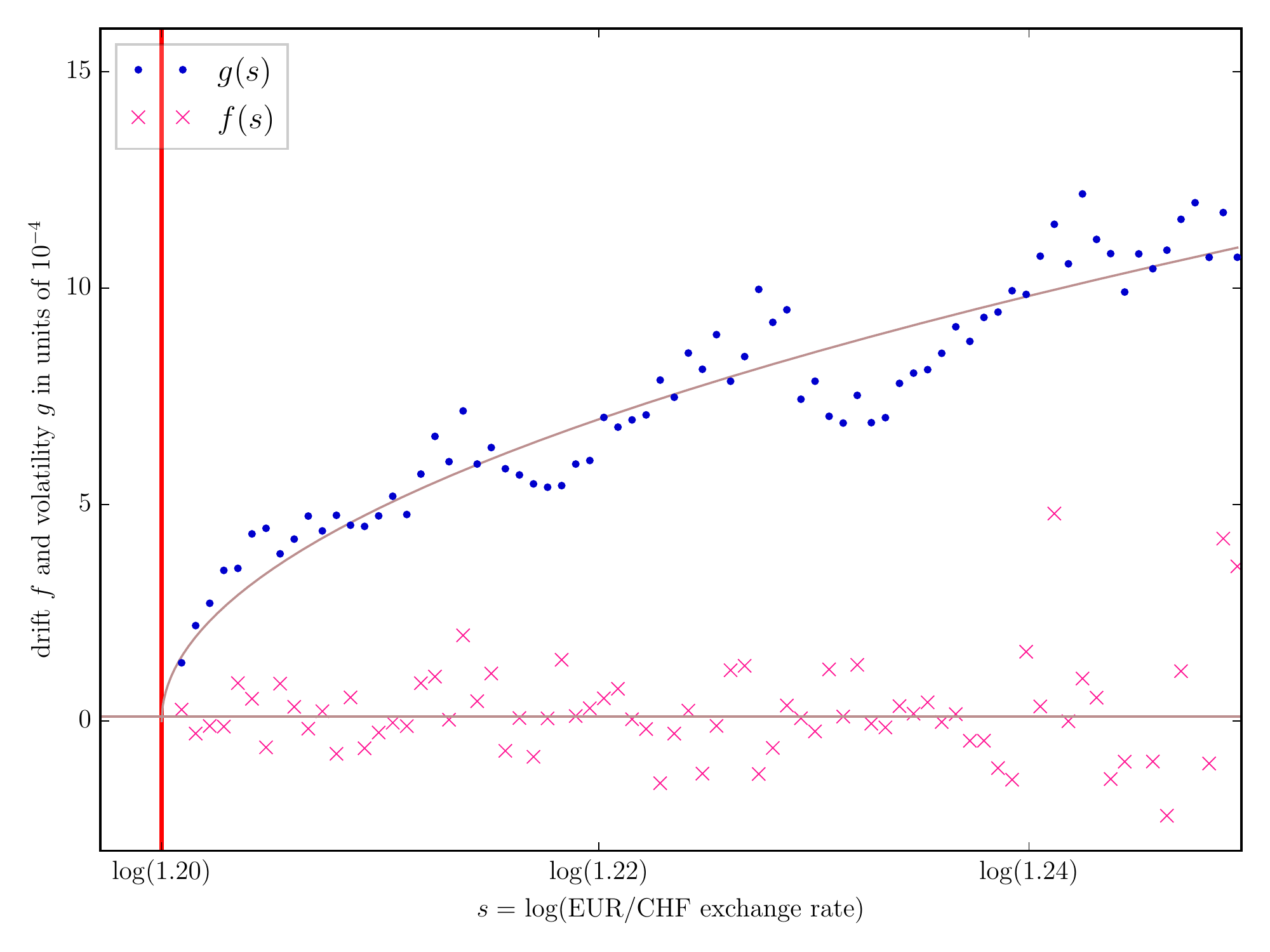}
	\caption{We show the parameter free estimate of drift $f(s)$ and volatility $g(s)$ obtained 
	from EUR/CHF exchange rate data between September 2011 and January 2015.} 
	\label{fig:f_and_g}
\end{figure}

We test the hypothesis \eqref{eq:physical_SDE} by extracting the terms $f$ and $g$
directly from the empirical data, using the definition \cite{Risken1996}
\begin{align}
f(s,t) &\equiv \lim_{\tau \to 0} \frac{1}{\tau} \mathbb{E} \left[ s(t+\tau) - s(t) \right]  \\
g(s,t) &\equiv \sqrt{\lim_{\tau \to 0} \frac{1}{\tau} \mathbb{E} \left[ \left( s(t+\tau) - s(t) \right)^2 \right]}
\end{align}
with $\mathbb{E} \left[ \cdot \right]$ the theoretical expectation operator. Assume that we are given 
a discrete time series consisting of $N$ data points $s_1, s_2, \ldots, s_N$, which 
is a discrete realisation of a stochastic process $\left\{ s(t) \right\}_{t \geqslant 0}$ (for instance a dataset of historical exchange rates). 
The temporal distance between two succeeding 
data points $s_i$ and $s_{i+1}$ is equal to $\tau$ ($0 < \tau \ll 1$) and assumed independent of $i$. 
Under the additional assumption that the
process is stationary,  $f(s,t) = f(s), g(s,t) = g(s)$, a parameter-free approach to extract $f$ and $g$ directly from this time series
is obtained by slicing up the value range $\left[ \min_i s_i, \max_i s_i \right]$ of the time series into $K$ bins $B_\ell, ~ \ell = 1, \ldots, K$ 
and approximating $f$ and $g$ in each bin according to \cite{Friedrich2000}
\begin{align}
	f \left( s^\ell \right) &\approx \frac{1}{\tau} \sum_{i \in B_\ell} \left( s_{i+1} - s_i \right) \label{eq:f_mean} \\ 
	g  \left( s^\ell \right) &\approx \sqrt{ \frac{1}{\tau} \sum_{i \in B_\ell}  \left(s_{i+1} - s_i  \right)^2}~, \label{eq:g_mean}
\end{align}
where $s^\ell$ denotes the mid point of the $\ell$-th bin and the 
summation is meant over all data points $s_i$ that lie in the $\ell$-th bin. 
For our application, we download tick by tick data of the EUR/CHF exchange rate, which is then coarse-grained to equally spaced 
time stamps of $10$ seconds ($\tau =1/360$ hours) by taking the median. 
The result is shown in figure \ref{fig:f_and_g} for $K = 100$ bins.
We have verified that the results are robust to different choices of the number of bins $K$ ranging at least from  $20$ to $140$ as well as with respect to sub-sampling at 
multiples of the initial time scale $\tau$ \cite{Sura2002}, confirming that the procedure \eqref{eq:f_mean} and \eqref{eq:g_mean} attains a reasonable linear convergence. 

Remarkably, we find that $f$ is essentially constant (and close to $0$), in complete contradiction with the constrained random walk entropic argument:
there is no entropic or other potential-derived force acting on the particle. The second interesting observation is that it is $g$ that exhibits a non 
trivial $s$ dependence. It turns out that this non trivial behavior of $g$ is intrinsic to the target zone regime. We have applied the algorithm 
of Friedrich et al. \cite{Friedrich2000} to EUR/CHF exchange rate data before September 2011. 
In the period preceding the committed action of the Swiss National Bank, 
$g$ remains approximately constant over a large range of values, thus recovering the standard GBM 
model. The corresponding figure is shown in \ref{sec:other_regimes}.

\section{Target zone modeling II: An economist's approach}
\label{sec:Krugman}

The empirical results from figure \ref{fig:f_and_g} clearly rejects the simple naive
physical model \eqref{eq:physical_SDE}. Before we go on and look for answers in a
more sophisticated physical model, let us consider first what is expected
from a purely economical point of view. 

\subsection{The no-arbitrage condition}

The fact that $f(s)$ is essentially zero for all $s$ reveals an important difference between
physical and economical Brownian motion. 
The exchange rate fluctuations are not due to unconcious random actions 
as would be the myriads of collisions of fluid molecules on a Brownian particle but due to the decisions of
investors trying to extract profit from their investments. The aggregate result of this behavior
of extremely motivated and driven agents is the quasi-absence of arbitrage, namely the 
impossibility to extract an excess return. The no-arbitrage condition is one of the
organising principles of financial mathematics and is expressed in general by the condition that
the process $s(t)$ obeying (\ref{eq:basic_SDE}) should be a martingale \cite{Samuelson1965}. 
In a risk neutral framework (which means that investors do not require
additional return for being exposed to risks), this translates mechanically into the condition of zero drift $f(s)=0$. 
In the presence of risk aversion, small values of $f(s)$ are present to remunerate the investors
from their expositions to the risks associated with the fluctuating prices. 
If there was a well-defined, significant drift, knowledge thereof could immediately be translated
into sure gains. 

To illustrate that this would be the case in our simplified physical model \eqref{eq:physical_SDE}, we
simulated synthetic time series with the generating process \eqref{eq:physical_SDE}, which has a non-zero 
$f(s)$, with  parameters chosen to match the empirical volatility. We used the simple strategy of 
selling (resp. buying) the euro and buying (resp. selling) the Swiss franc whenever $s > s_{\rm eq}$
(resp. $s < s_{\rm eq}$). Including typical transaction costs between 1 and 2 pips (1 pip $=0.0001$ is
approximately equal to the bid-ask spread of the real EUR/CHF tick data from Sept. 6, 2011 to January 14, 2015), we find 
this strategy to deliver extremely high, two-digit annualised Sharpe ratios (as a benchmark, it is 
typical for mutual funds, hedge-funds and the market portfolio itself to deliver performances with 
Sharpe ratios less than $1$, and often much less then $1$). This clearly
illustrates that the process \eqref{eq:physical_SDE} would lead to exchange rates that can be forecasted, which
would ``leave enormous amount of money on the table''.
It is thus completely unrealistic from a financial view point.

\subsection{Brief summary of Krugman's model}

\begin{figure}[!htb]
	\centering
	\includegraphics[width = 0.4\textwidth]{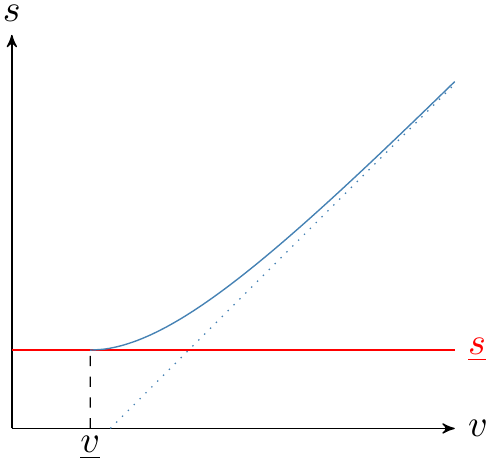}
	\caption{The plain line denotes the exchange rate $s$ (how many units of foreign currency we can buy with one unit of
	domestic currency) as a function of the exogenous velocity shocks $v$, see equation \eqref{eq:Krugman_solution}. 
	Decreasing $v$ indicates that the economic and geopolitical environment is such that the domestic currency is 
	gaining in value relative to the foreign one. The dotted line denotes the relation between $s$ and $v$ in the absence of 
	a target zone, i.e. when the monetary supply $m$ is held constant. In the limit $v \gg \underline{v}$ the two curves coincide because
	the presence of the barrier is not felt. As $v$ approaches $\underline{v}$, the central bank increases $m$, thus keeping $s$ artificially 
	above or exactly at $\underline{s}$.}
	\label{fig:Krugman_solution}
\end{figure}

The work of Krugman \cite{Krugman1991} turns out to be the reference
of a large part of the economic target zone literature. According to Krugman, the constrained exchange rate $s$ can be described as
\begin{equation}
	s = m + v + \gamma \frac{ \mathbb{E} \left[ \mathrm{d}s \right]}{\mathrm{d}t}.
	\label{eq:Krugman_start}
\end{equation}
By $m$, we denote the (logarithm of) the money supply. As long as $s$ is above the lower boundary $\underline{s}$, $m$ is 
supposed to be held constant. Once $s$
touches $\underline{s}$, the central bank (here the SNB) is supposed to increase the money supply, thus weakening the domestic currency (CHF) relative to the foreign one (EUR),
which means that $s$ is pushed away from the lower boundary. By $v$, we denote the (logarithm of) exogenous velocity shocks, i.e. influences on the exchange rate
coming from the economic and politic environment that cannot be controlled by the national bank. It
is assumed that $v$ follows a standard Brownian motion 
\begin{equation}
	\mathrm{d}v = \sigma \mathrm{d}W_t ~~~ (\sigma > 0).
	\label{eq:dv}
\end{equation}
The last ingredient to Krugman's model is the expected change in $s$, $\mathbb{E} \left[ \mathrm{d}s \right] / \mathrm{d}t$. 
It is this term that makes all the difference between the naive physical model and Krugman's economical model. 
The reasoning behind this term is that, as $s$ approaches $\underline{s}$ from 
above, market participants anticipate the central bank's intervention and act accordingly. 
This is different from the unconscious physical particle that, as long as there is no contact (either direct or mitigated
through the fluid), behaves as if there was no wall. \\ 
The constant $\gamma$ denotes the semi-elasticity of the exchange rate
with respect to the instantaneous expected rate of currency depreciation. 
Equation \eqref{eq:Krugman_start} can be solved with basic stochastic calculus, see \cite{Krugman1991}. The result reads
\begin{equation}
	s = m + v + A e^{- \rho v}
	\label{eq:Krugman_solution}
\end{equation}
where $\rho = \sqrt{2 / \gamma  \sigma^2}$. Denote by $\underline{v}$ the unique value of $v$ 
at which $s(v = \underline{v}) = \underline{s}$ (for fixed $m$, understood in the limit $v \downarrow \underline{v}$). Then, the constant $A$
is determined uniquely by demanding that the derivative of $s$ as a function of $v$ vanishes at $\underline{v}$, 
\begin{equation}
	\left. \frac{\mathrm{d}s}{\mathrm{d}v} \right|_{v = \underline{v}} = 0.
\end{equation}
This condition is rooted in a no-arbitrage argument known in option pricing as smooth pasting \citep{Dixit1993}.
The final result is depicted in figure \ref{fig:Krugman_solution}. 

\subsection{Assumptions of Krugman's model}

Krugman's target zone model is based on two crucial assumptions: First, the target zone is perfectly credible. This means that market participants
belief at every time that the central bank will stick to its announced target zone. 
Second, the interventions by the central bank are marginal, meaning
the monetary supply is held constant as long as $s$ is within the target zone band. Only when $s$ touches $\underline{s}$, the monetary 
supply is increased, just sufficiently to keep $s$ at $\underline{s}$. 
These assumptions have been investigated specifically for the EUR/CHF exchange rate between 2011 and 2015 in \cite{StuderSuter2014}. It is found that the two assumptions hold sufficiently
well so that Krugman's model can be applied. This sets the EUR/CHF target zone apart from many earlier empirical studies in which Krugman's model
was already challenged on the basis of its assumptions. We refer to \cite{Svensson1992,Sarno2003} for detailed reviews. 

\subsection{Drift and volatility in Krugman's model}

By applying It\^o's lemma to \eqref{eq:Krugman_solution}, we derive the following
drift $f$ and volatility $g$ in the Krugman framework:
\begin{align}
	f(v)	&=	\frac{1}{2} A \sigma^2 \rho^2 e^{- \rho v} \label{eq:Krugman_drift_v} \\
	g(v)	&=	\sigma - \sigma A \rho e^{- \rho v}. \label{eq:Krugman_volatility_v} 
\end{align}
For practical purposes, working with \eqref{eq:Krugman_drift_v} and \eqref{eq:Krugman_volatility_v} is
cumbersome because $v$ cannot be measured but only estimated \cite{Flood1991}. Nevertheless, testing
directly the non-linear $s(v)$ relation \eqref{eq:Krugman_solution} by estimating $v$ is the method 
that has been widely applied in the empirical literature.
The reported results have then either rejected Krugman's target zone model entirely or have shown only 
a very noisy evidence for \eqref{eq:Krugman_solution}. We refer again to  \cite{Svensson1992,Sarno2003}
for a broad overview and to \cite{StuderSuter2014} for EUR/CHF specific results. 

Our strategy is different. Instead of relying on $v$, we invert the $s(v)$ relation \eqref{eq:Krugman_solution}
locally to lowest order in $v-\underline{v}$ (it is easy to see that \eqref{eq:Krugman_solution} has
a well-defined, global inverse $v(s)$ that, however, has no analytical closed form expression). 
For $s$ close to $\underline{s}$, a second-order Taylor expansion gives
\begin{equation}
	s(f) \approx  \underline{s} + \frac{1}{2} \rho \left( f - \underline{f} \right)^2.
	\label{eq:s(f)}
\end{equation}
Inverting \eqref{eq:s(f)} yields $f(s)$ and plugging this into \eqref{eq:Krugman_drift_v} and \eqref{eq:Krugman_volatility_v},
the following expressions for drift and volatility are found: 
\begin{align}
	f(s) 	&= \alpha  \label{eq:Krugman_drift} \\
	g(s)	&= \beta \sqrt{s - \underline{s}}  \label{eq:Krugman_volatility} 
\end{align}
where $\alpha = \sigma \left/\sqrt{2 \gamma} \right., ~\beta = 2^{3/4} \sqrt{\sigma} \left/ \gamma^{1/4} \right.$. In particular,
we note that $\sqrt{\alpha} / \beta = 1/2$.  
There are higher order terms leading to corrections to \eqref{eq:Krugman_drift} and \eqref{eq:Krugman_volatility}. It is 
easy to check that, for our data where $s < \log (1.26)$, these corrections are negligible. 

Comparing \eqref{eq:Krugman_drift} and
\eqref{eq:Krugman_volatility} with figure \ref{fig:f_and_g}, one can check that the data conform very well to Krugman's theory. 
For the volatility, we can apply a one parameter least-squares fit which determines $\beta = \left( 5.42 \pm 0.06 \right) \cdot 10^{-3}$. Another least-squares fit determines $\alpha =  \left( 1.40 \pm 0.8 \right) \cdot 10^{-5}$. Basic error propagation calculations yield
$\sqrt{\alpha} / \beta = 0.68 \pm 0.22$. Despite the relatively large fluctuations for $s \gtrsim \log(1.24)$, the data 
agrees with the theoretical value $1/2$ within one standard deviation. Ignoring the large fluctuations around  $s \gtrsim \log(1.24)$
leads to even better correspondence between data and theory. We have also applied a maximum likelihood ratio 
test of nested hypotheses, which consists in 
comparing the hypothesis \eqref{eq:Krugman_volatility} to the more general $g(s) = \beta (s - \underline{s})^\mu$
with variable $\mu$. We find a $p$-value of $0.74 $ (above the standard confidence bound $0.05$), 
which means that the extended model with fitted exponent $\mu$ is not necessary and that 
expression (\ref{eq:Krugman_volatility}) with the fixed $\mu=1/2$ is sufficient to describe the data.
This confirms that Krugman's target zone model provides a suitable
description of the constrained EUR/CHF exchange rate.  

\begin{figure}[!htb]
	\centering
	\includegraphics[width = 0.5\textwidth]{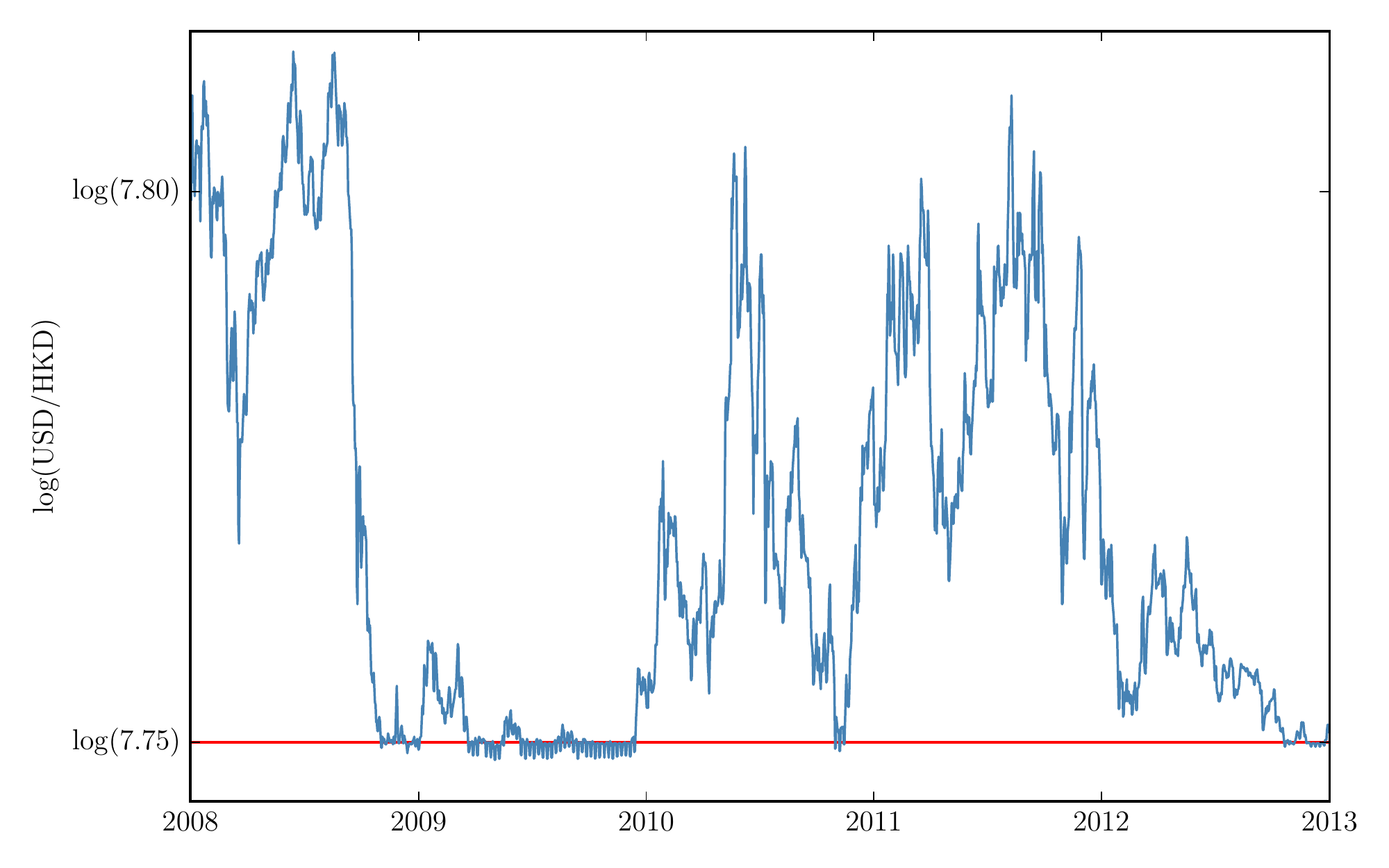}
	\caption{USD/HKD exchange rate from January 2008 to December 2012. The red line denotes the lower cap of the USD/HKD target zone.} 
	\label{fig:USD_HKD}
\end{figure}

We stress that this result is novel also from the perspective of pure economical research. First, by inverting
locally the relationship between $s$ and $v$, we have derived a way by which the model can be tested directly, 
without the need of estimating the (usually unknown and unknowable) fundamental value of the exchange rate. 
Second, the theoretically celebrated Krugman model has been mostly rejected by previous empirical studies. 
Although target zones are not a new concept and have been particularly popular in the European Monetary System (EMS) 
from the 1970s to the 1990s, never has there been such a consistent pressure keeping the 
exchange rate remarkably close to the target zone boundary over large periods of time, as was observed for
the EUR/CHF target zone.  Another example of 
a recently introduced target zone is the one of the US dollar/Hong-Kong dollar (USD/HKD) exchange rate implemented
by the Hong-Kong Monetary Authority (HKMA) in 2005 (figure \ref{fig:USD_HKD}).
It is easy to see that here, unlike for the EUR/CHF target zone, 
the pressure is not as consistent and the barrier not as vehemently defended (there are small fluctuations around the target zone limit). This disqualifies many other target zones 
already due to the restrictive assumptions of Krugman's target zone model and highlights the special case of the EUR/CHF target zone. 
Thus, we have shown that Krugman's model holds after all, but not for arbitrary target zones, 
but only under extreme and sustained pressure (in this case associated with the European crisis) that pushes continuously the exchange
rate very close to the boundary of the target zone, which is rigorously defended by the central bank.

\section{Target zone modeling III: Hindered diffusion}
\label{sec:hindered_diffusion}

We have shown in the previous section that the initial naive physical model failed
to account to for the action anticipating traders. In broader terms, we can classify 
this finding as ``the presence of the wall must be felt even away from the wall" 
at all times and for all random walk realisations. We use this insight to develop a second, 
more elaborate physical model. We note that the physical Brownian particle can receive
the information ``there is a wall" not only through direct contact but also through the fluid. 

\begin{figure}[!htb]
	\centering
	\includegraphics[width = 0.5\textwidth]{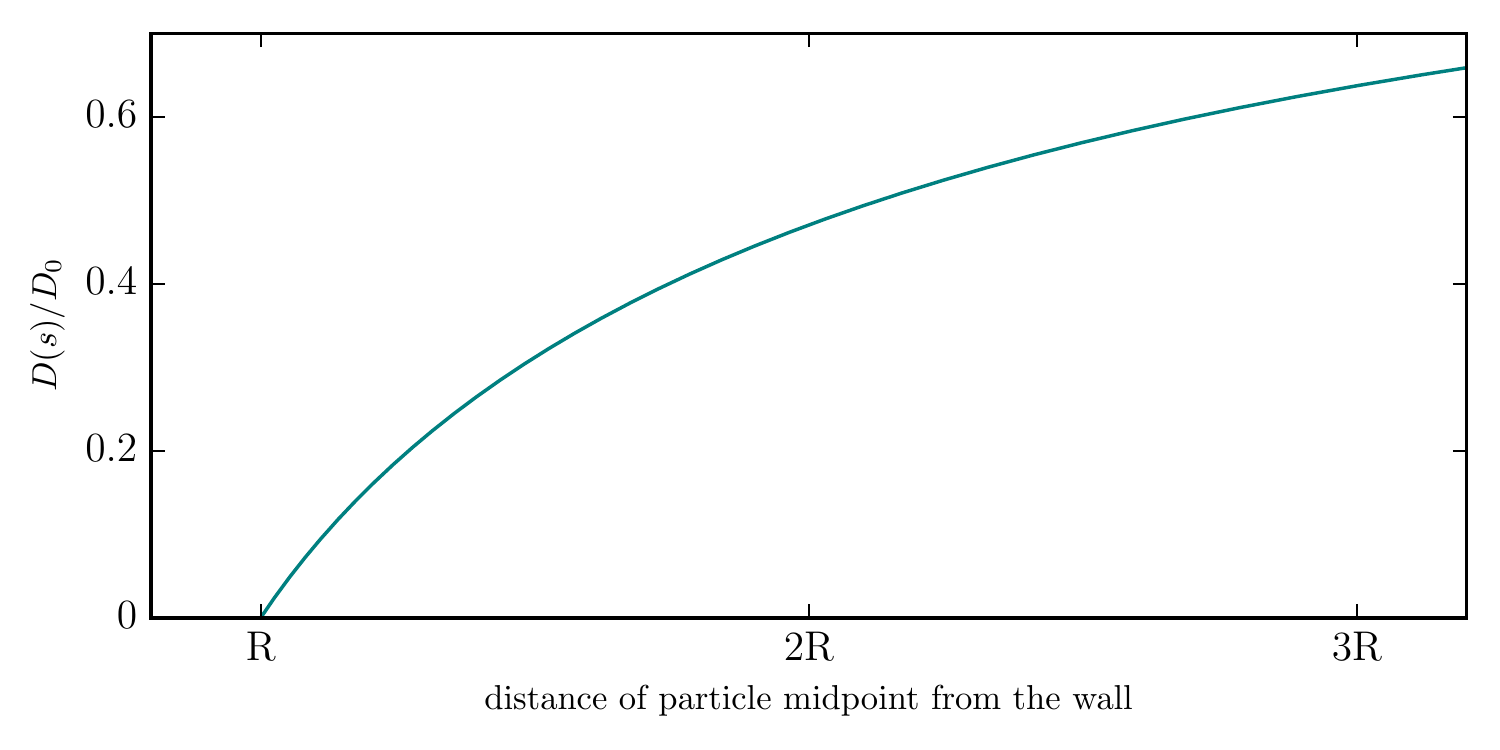}
	\caption{Physical diffusion coefficient as a function of particle distance from the wall.
		To first order and close to the wall, $D(s)$ is a linear function of $s-\underline{s}$.} 
	\label{fig:diffusion_coefficient}
\end{figure}

Consider a physical Brownian particle in a fluid. The presence of a wall leads to a modification of the hydrodynamic flow
of the molecules trapped between the wall and the Brownian particle.
The closer the Brownian particle to the wall, the thinner the lubrification layer between them
and the more hindered is the diffusion of the Brownian particle. 
In physics, it is more common to work with the diffusion coefficient $D(s)$ 
which is related to our volatility via $g = \sqrt{2D}$. In the bulk of a fluid (where the wall is not felt),
the diffusion coefficient $D$ is a constant $D_0$. The Einstein-Stokes 
equation predicts for a spherical particle with radius $R$
\begin{equation}
	D_0 = \frac{k_B T}{6 \pi \nu R} 
	\label{eq:D0}
\end{equation}
with $k_B$ the Boltzmann constant, $T$ the temperature and $\nu$ the
viscosity of the fluid. In presence of a wall at $s = \underline{s}$, Brenner  \cite{Brenner1961} showed that the diffusion coefficient must be 
modified by $D(s) = D_0 / \lambda$
where $\lambda$ depends in a complicated, non-linear way on the ratio of
$s-\underline{s}$ and $R$ (equation (2.19) in \cite{Brenner1961}). The result is depicted in figure \ref{fig:diffusion_coefficient}
and shows that the presence of a barrier translates into
the decrease due to hydrodynamic forces of the diffusion coefficient of the Brownian particle upon its approach to the wall.
An approximation of this result had already been found decades before by Lorentz \cite{Lorentz1907} who predicted 
\begin{equation}
	\lambda \sim 1 + \frac{9}{8} \frac{R}{s-\underline{s}}
	\label{eq:Lorentz}
\end{equation}
from which we find, to first order, $D(s) \sim (s-\underline{s})$. 
It follows immediately that, close to the wall, the volatility ($g = \sqrt{2D}$) 
of the particle increases like the square-root of $s-\underline{s}$, 
in correspondence with the model \eqref{eq:Krugman_volatility} from finance 
(it is clear that the constant pre-factor depends on physical quantities that are unrelated to the economical analogy). 
Hence, we have shown that there is an analogy between physical and economical Brownian motion after all. 
Had we relied from the beginning on the correct physical picture of hindered diffusion, we could have predicted the
behavior of the exchange rate close to the barrier solely on the basis of physical theories that were known
already long before Krugman's model (up to 
a scale factor, discussed in the next section). 

The analogy that we have found here extends a previous result \cite{Yura2014} in which it has been shown 
that the GBM model of financial price fluctuations is deeply anchored in 
the physics-finance analogy of a colloidal Brownian particle embedded in a fluid of molecules
as shown in figure \ref{fig:particle_in_potential} (omitting the previously shown incorrect potentials),
where the surrounding molecules reflect the structure of the underlying order book.  Here, we have
shown that this analogy holds even if the order book fluid is restricted from one side by an upper or lower bound.

\section{Equilibrium in physics and in economics}
\label{sec:equilibrium}

In absence of any external force, what is the stochastic process that describes a physical Brownian particle? Naively, 
one is led to propose $ds/dt = g(s) \cdot \eta(t)$. However, this implies not only that we are working in It\^o's interpretation
of stochastic calculus, but can furthermore be shown to be inconsistent with convergence towards thermal equilibrium. 
For a system at equilibrium, the probability density $p(s,t)$ must have a steady state solution 
with the canonical form $p(s) \sim \exp(- \mathcal{H}/ k_B T)$ with $\mathcal{H}$ the Hamiltonian of the
system. If we insist on working in It\^o's interpretation as is customary in finance to ensure
causality of financial strategies, one must include an additional drift term $g(s) \frac{dg(s)}{ds}$ to the stochastic differential 
equation in order to be consistent with the physical steady state distribution
(see  section 2.2.3 of \cite{Sornette2003} and \cite{Lau2007} for derivations). From \eqref{eq:Krugman_volatility}, 
we then derive the following stochastic equation for a Brownian particle close the a wall and in absence of external forces: 
\begin{equation}
	\frac{\mathrm{d}s}{\mathrm{d}t} = g(s) \frac{\mathrm{d} g(s)}{\mathrm{d} s} + g(s) \cdot \eta(t) = \frac{\beta^2}{2} + \beta \sqrt{s-\underline{s}} \cdot \eta(t). 
	\label{eq:sde_Brownian}
\end{equation}
Remarkably, the square-root shaped volatility is exactly the function that induces a constant positive drift in agreement with
Krugman's prediction \eqref{eq:Krugman_drift}. From a purely physical perspective, this result \eqref{eq:sde_Brownian}
has another interesting implication. It reveals the special role played by the linearly increasing diffusion coefficient. 
It can be shown that a locally linear diffusion coefficient is the only physically sensible choice. Since this result is not 
the main concern of our paper, we refer the interested reader to \ref{sec:linear_diffusion} for its derivation.

The correspondence between physical hindered diffusion and Krugman's target zone model is therefore only semi-quantitative 
in the sense that here 
$\sqrt{\text{"drift term"}} / \beta = 1/\sqrt{2}$. For Krugman, on the other hand, we have derived $\sqrt{\text{"drift term"}} / \beta = 1/2$, thus revealing a key difference between Krugman's constant drift term 
and the one resulting from a noise-induced drift of the form \eqref{eq:sde_Brownian}. 
We attribute this difference of the numerical values of $\sqrt{\text{"drift term"}} / \beta$ to 
the global condition of thermal equilibrium $p(s) \sim \exp(- \mathcal{H}/ k_B T)$, which is 
absent in finance. L\'evy and Roll \cite{Levy2010} have recently proposed to 
impose the constraint that the global market portfolio is mean-variance efficient, i.e, that it
obeys the predictions of the Capital Asset Pricing Model (CAPM). This global condition
can be shown to lead to a reassessment and an improved estimation of the expected
returns of the stocks constituting the global market \cite{Nietal2011}.
But it is not known what could be other consequences, in particular in exchange rate dynamics.
Indeed, in finance, the existence of an economic equilibrium distribution
similar to Boltzmann, and its relation to detailed balance is highly debated 
and far from trivial. We refer to  \cite{Fiebig2015,Farmer2009} for recent discussions of this topic and 
to \cite{Sornette2014,Bouchaud2001,Farmer2008,Zhang1999,Derman2004} 
for further details on the interplay and coevolution of physics and economics in general.

\section{Conclusions}
\label{sec:conclusions}

This paper has served two purposes. From a technical perspective, 
we have shown that the constrained EUR/CHF exchange rate is well-described
by Krugman's target zone model \cite{Krugman1991}, which incorporates the traders' expectations 
as a fundamental ingredient into its equations. By describing the exchange rate as a colloidal Brownian particle embedded in an 
``order book fluid", we could show furthermore 
that there is a formal analogy to the physical hindered
diffusion problem in the sense that both systems can be described by the same stochastic differential equation. 
This provides novel empirical support for the recently introduced model of a
``financial Brownian particle in a layered order book fluid'' \cite{Yura2014}, which generalises the standard
random walk model of financial price fluctuations.  

From a didactical perspective, we have given a complete example of how economic models
can be motivated by ideas and mathematical tools from physics and vice versa. 
We have also pointed out a fundamental
difference between physical and economic hindered diffusion. In physics, 
we have an additional constraint in terms of a thermal equilibrium based on
detailed balance. In finance, the existence of such a global equilibrium is a priori not clear
and must be investigated further. It would not be unsurprising, if further motivations or analogies
can be found from the study of physical out-of-equilibrium systems.

\section*{Acknowledgment}
\label{sec:acknowledgment}

The research performed by one of us (SL), which is reported in this article,
was conducted at the Future Resilient Systems at the Singapore-ETH Centre (SEC). 
The SEC was established as a collaboration between ETH Zurich and National Research Foundation (NRF) 
Singapore (FI 370074011) under the auspices of the NRF's Campus for Research Excellence and Technological 
Enterprise (CREATE) programme.

\bibliographystyle{unsrt}
\bibliography{PhysRevE}

\appendix
\section{Target zone dependence}
\label{sec:other_regimes}

\begin{figure}[!htb]
	\centering
	\includegraphics[width =  0.5 \textwidth]{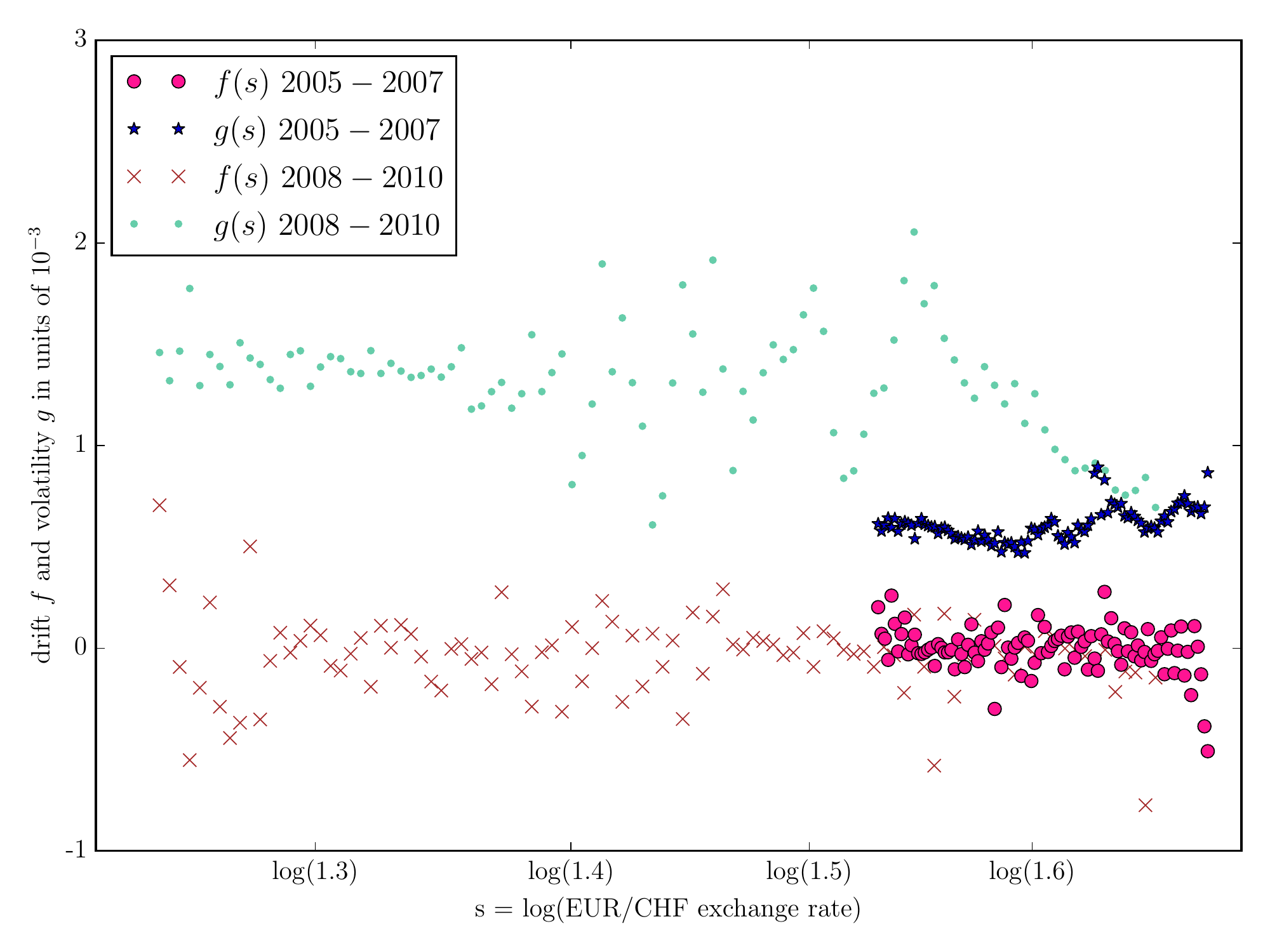}
	\caption{Approximation of drift $f$ and volatility $g$ using 10 seconds data of the EUR/CHF exchange rate from 2005 to 2007. 
	This figure is obtained by applying the algorithm by Friedrich et al. \cite{Friedrich2000}. 
	In contrast to the target zone regime, we observe that here $g$ is roughly constant.}
	\label{fig:2005_2010} 
\end{figure}

For the edifice of this paper, it is vital to show that the square-root shaped 
volatility is intrinsic to the target zone regime from September 2011 to January 2015. 
Indeed, applying the algorithm of Friedrich et al. \cite{Friedrich2000} to 
EUR/CHF exchange rate data ranging from 2005 to 2007 and from 2008 to 2010 (figure \ref{fig:2005_2010})
shows that $g$ is roughly constant over a large regime of values. 
We have chosen data ranging over periods of three years in order to
have approximately the same amount of data points as during the target zone regime. 

\section{Diffusion close to a wall}
\label{sec:linear_diffusion}

Working with It\^o's interpretation of stochastic calculus, it can be shown 
(section 2.2.3 of \cite{Sornette2003} and \cite{Lau2007}) that a Brownian particle with general diffusion coefficient
$D(s) = g(s)^2/2$ and in absence of any external forces is
described by the stochastic differential equation
\begin{equation}
	\frac{ds}{dt} = g(s) \frac{d g(s)}{ds}  + g(s) \cdot \eta(t).
	\label{eq:Brownian_SDE_Lau2007}
\end{equation}
We want to determine the volatility $g(s)$ of a Brownian 
particle at position $s$ close to a wall located at $s = \underline{s}$. 
This problem was first solved in an exact (but fairly complicated) manner by Brenner \cite{Brenner1961}, 
stating that the bulk diffusion coefficient \eqref{eq:D0} must be replaced by $D_0/\lambda$. 
Without loss of generality, we set now $\underline{s} = 0$. From Lorentz' approximate result
\eqref{eq:Lorentz}, we infer that close to the wall, $D(s) = D_0/\lambda$ is, to first order, linear in $s$. 

In this appendix, we want to give a less rigorous but simple heuristic derivation of this result. 
What is nice about our derivation is that no detailed knowledge about hydrodynamic interactions is required. 
We make the fairly general approximation that, close to the wall,
$g(s) = \beta s^\gamma$ for some $\gamma > 0$ (it is easy to see that 
$\lim_{s\downarrow 0} D(s) = 0$ is a necessary condition). 
Plugging this
into \eqref{eq:Brownian_SDE_Lau2007} gives
\begin{equation}
	\frac{ds}{dt} = \beta^2 \gamma s^{2 \gamma -1} + \beta s^\gamma \cdot \eta(t).
	\label{eq:ds_Brownian_Ito}
\end{equation}
In the limit $s \downarrow 0$, we can distinguish three cases: 
\[
\text{the drift } g(s) \frac{dg(s)}{ds} ~~
\left\{	\begin{array}{cc} \text{diverges} & \text{if } \gamma < 1/2,\\
				\text{is constant}& \text{if } \gamma = 1/2, \\
				\text{vanishes} & \text{if } \gamma > 1/2. 
	\end{array} \right.
\]
If $\gamma < 1/2$, the particle will be repelled with infinite force and can never touch the wall. Furthermore, 
placing initially the particle at the wall is ill-defined.  
If $\gamma > 1/2$, the particle, once it has reached the wall, 
will stay there forever (more precisely, it can be shown that a particle starting from $s > 0$ can never exactly reach the wall, but
approach it arbitrarily close \cite{Gardiner1985}). Also, a particle placed at the the wall will simply stay there forever.
We deduce that $\gamma = 1/2$, and hence $D(s) \sim s$ is the only 
physically reasonable choice. In this case, a particle starting from $s \geqslant 0$ has non-zero probability to reach the
boundary in finite time, upon which it will be repelled. 
These arguments can be formalised by solving analytically the Fokker-Planck equation 
corresponding to \eqref{eq:ds_Brownian_Ito} in terms of an eigenfunction expansion \cite{Risken1996,Ruseckas2010}. 

\end{document}